\documentclass[conference,a4paper]{APSIPA2021}
\usepackage{amsmath}
\usepackage{graphicx}
\usepackage{threeparttable}
\usepackage[caption=false]{subfig}
\usepackage[dvips]{epsfig}
\usepackage{booktabs}
\usepackage{multirow}
\usepackage{array}
\usepackage{url}
\usepackage{geometry}
\usepackage{fancyhdr}

\fancypagestyle{firststyle}{
  \fancyhf{}
  \fancyhead[C]{2023 Asia Pacific Signal and Information Processing Association Annual Summit and Conference (APSIPA ASC)}
}

\begin{document}

\title{An Analysis of Personalized Speech Recognition System Development for the Deaf and Hard-of-Hearing}

\author{
\authorblockN{
Lester Phillip Violeta and
Tomoki Toda
}

\authorblockA{
Nagoya University, Japan \\
E-mail: violeta.lesterphillip@g.sp.m.is.nagoya-u.ac.jp}
}

\maketitle
\thispagestyle{firststyle}
\pagestyle{fancy}

\begin{abstract}
Deaf or hard-of-hearing (DHH) speakers typically have atypical speech caused by deafness. With the growing support of speech-based devices and software applications, more work needs to be done to make these devices inclusive to everyone. To do so, we analyze the use of openly-available automatic speech recognition (ASR) tools with a DHH Japanese speaker dataset. As these out-of-the-box ASR models typically do not perform well on DHH speech, we provide a thorough analysis of creating personalized ASR systems. We collected a large DHH speaker dataset of four speakers totaling around 28.05 hours and thoroughly analyzed the performance of different training frameworks by varying the training data sizes. Our findings show that 1000 utterances (or 1-2 hours) from a target speaker can already significantly improve the model performance with minimal amount of work needed, thus we recommend researchers to collect at least 1000 utterances to make an efficient personalized ASR system. In cases where 1000 utterances is difficult to collect, we also discover significant improvements in using previously proposed data augmentation techniques such as intermediate fine-tuning when only 200 utterances are available. 
\end{abstract}

\section{Introduction}
Deaf or hard of hearing (DHH) people typically have their development of speech stunted due to hearing impairedness. This results in difficult-to-understand speech \cite{dhh_speech}, or commonly referred to as atypical speech, which makes daily conversations a tedious task for them. As communication is one of the essential aspects of being human, there is a need to develop more systems to support DHH speakers and provide them with more communication alternatives.

Recent advancements in artificial intelligence (AI) has given way for computers to assist humans in their everyday life. One particular field with substantial improvements in the past decade is automatic speech recognition (ASR) \cite{asr1, asr2, asr3}, the conversion of speech into its corresponding text labels. Owing to neural network-based approaches and the availability of large-scale datasets, several systems have been proposed that could transcribe text at an astounding accuracy rate. Smart devices, in particular voice assistants, have used this technology as a means of communication between users' speech and a control system to execute the said command, enabling several conveniences such as controlling appliances around the house. Thus, with the growing popularity of ASR, several devices and software applications now support using speech to navigate through the device or the application. For DHH users, using ASR to transcribe their spoken speech into text becomes available as an alternative communication option. With the huge benefits and conveniences that ASR-based smart devices bring, improving and developing personalized ASR systems for DHH speakers can allow them a communication alternative and be able to use these devices as an assistant in everyday life. 

Although ASR systems have been successful with typical speech, it has struggled with transcribing DHH speech due to the difference and high variability in the DHH speech patterns compared to typical speech. %As seen in Figure \ref{fig:spec}, 
The deaf speaker usually cannot control the high-frequency components of speech \cite{deaf_perception} due to their hearing conditions, making how other humans and ASR models perceive their speech entirely different from typical speakers. Consequently, this means that although ASR models are usually trained on large-scale typical speech (usually a minimum of 1000 hours), the success of these ASR models is also limited to just typical speech. Replicating the same success on DHH speech would require training the ASR model on a large-scale DHH speaker dataset; however, recording at a large scale, due to strict ethics regulations and the effort it takes to setup recordings, makes this a difficult task to resolve. Thus, speakers with atypical speech have found these devices rather difficult to use \cite{inclusive_asr_deaf, inclusive_asr}. However, as mentioned, DHH users would be the ones who could benefit the most from ASR-based devices.

% \begin{figure}[ht]
%     \includegraphics[width=155mm]{img/spectrogram}
%     \caption{
%     Spectrogram differences of a typical speaker and a deaf speaker uttering the same sentences.
%     }
%     \label{fig:spec}
% \end{figure}

In this research, we provide an analysis of how to create personalized ASR systems for DHH speakers. Our goal is to provide a thorough analysis such that researchers would be able to understand the training size/performance tradeoff when collecting speech data, and how to make use of methodologies used in the literature to properly develop personalized ASR systems. Our previous work has shown the effectiveness of large-scale pretraining and using imperfect synthetic data to improve recognition accuracy in electrolaryngeal speech recognition \cite{el-intermediate-finetuning}; however, there is still some doubt on whether the previously proposed method works on different types of atypical speech or on larger dataset sizes. Thus, this extension of the analysis provides information on the effects of varying the training data sizes and how it affects the performance of each ASR method. Moreover, compared to other atypical datasets used in the literature, we collect a large dataset of four DHH speakers with 28.05 hours in total, allowing us a thorough examination of the dataset. %Our findings confirm that our previously proposed data augmentation techniques can significantly improve performance when data is limited to 200 utterances even in DHH speech; however, collecting as few as 1000 utterances from a speaker is sufficient enough to get huge performance boosts by just fine-tuning a pretrained ASR model. 
Our main findings are as follows:

\begin{itemize}
  \item We perform a thorough analysis with publicly available ASR resources on four different DHH Japanese speakers to make a personalized ASR system. We train personalized models and analyze different scenarios with different training data sizes and data augmentation techniques to give researchers an idea of what requirements are needed to train high-performing personalized ASR systems.
  \item Through our findings, we confirm that our previously proposed speech synthesis-based data augmentations indeed improve ASR performance in low-resourced scenarios (200 utterances) even on DHH speech datasets. However, using at least 1000 utterances of real recordings to fine-tune pretrained models already shows huge performance boosts in ASR performance, and thus recommend researchers collect around this number of utterances as it is the best cost/performance tradeoff point. 
  \item Finally, we emphasize the effectiveness of creating personalized ASR systems, as when conducting data augmentation, using the synthesized speech of the target speaker was always better than using real recordings from other speakers.
\end{itemize}

\section{Related literature}
Since the number of devices and software applications that support the use of speech as an input grows more and more each day, using ASR as a communication alternative is becoming more available as an option. However, research such as \cite{inclusive_asr_deaf, inclusive_asr} has investigated that for DHH and atypical speech, currently productionized or open-sourced ASR models have up to 60\% increase in word error rates (WER) compared to typical speech. Thus, although ASR-based technology has been gaining popularity and more support on devices, current models are still far from being inclusive to all people, and more investigations need to be done to resolve this issue.

Although not focused on DHH speech, several research works have investigated improving ASR performance for atypical speakers. %As collecting speech data from atypical speakers is a difficult task and there are only a limited number of publicly available atypical speech datasets, several papers have investigated resolving the low-resourced problem in atypical speech recognition. 
The most common method to resolve this is by pretraining the model with large-scale typical speech data, and subsequently fine-tuning on the target low-resourced atypical speech data \cite{project-euphonia, pretrain-dysarthric-shor, pretrain-dysarthric-outperforming, pretrain-pathological-ssl}. Through large-scale pretraining, the model can generalize to the target better by using these as the initialization model parameters, rather than training the model from randomized initialization model parameters. The larger the pretraining dataset used, the more generalized the model could be and thus more robust to different types of speech during fine-tuning.

Research such as \cite{pretrain-dysarthric-outperforming, pretrain-dysarthric-personalized} has shown that personalized ASR models will always outperform ASR models trained on other speakers' data. Although this work has shown that simply fine-tuning a pretrained model with target speakers can massively improve ASR performance, our previous research in \cite{el-intermediate-finetuning} shows that naively fine-tuning the pretrained model on the target speaker is not very effective in a low-resourced scenario, such as when there are only 200 utterances available for each speaker. To further improve performance, we proposed using an intermediate fine-tuning approach using imperfect synthetic speech by using text-to-speech (TTS) \cite{el-intermediate-finetuning}. The main point made here was that even low-quality synthetic speech of the target speaker (in this case, the electrolaryngeal speaker) could be utilized to improve ASR performance, contrary to the common intuition that data augmentations should always be of high quality. This has opened doors in improving low-resourced ASR, by requiring looser requirements to train ASR models of a target atypical speaker. However, more work has yet to be done to verify the effectiveness of this method on other types of atypical speakers (such as DHH) or if it would still work well in scenarios with a larger training data size as the experiments were only conducted using a small-scale electrolaryngeal speaker dataset.

\begin{figure}[t!]
    \includegraphics[width=90mm]{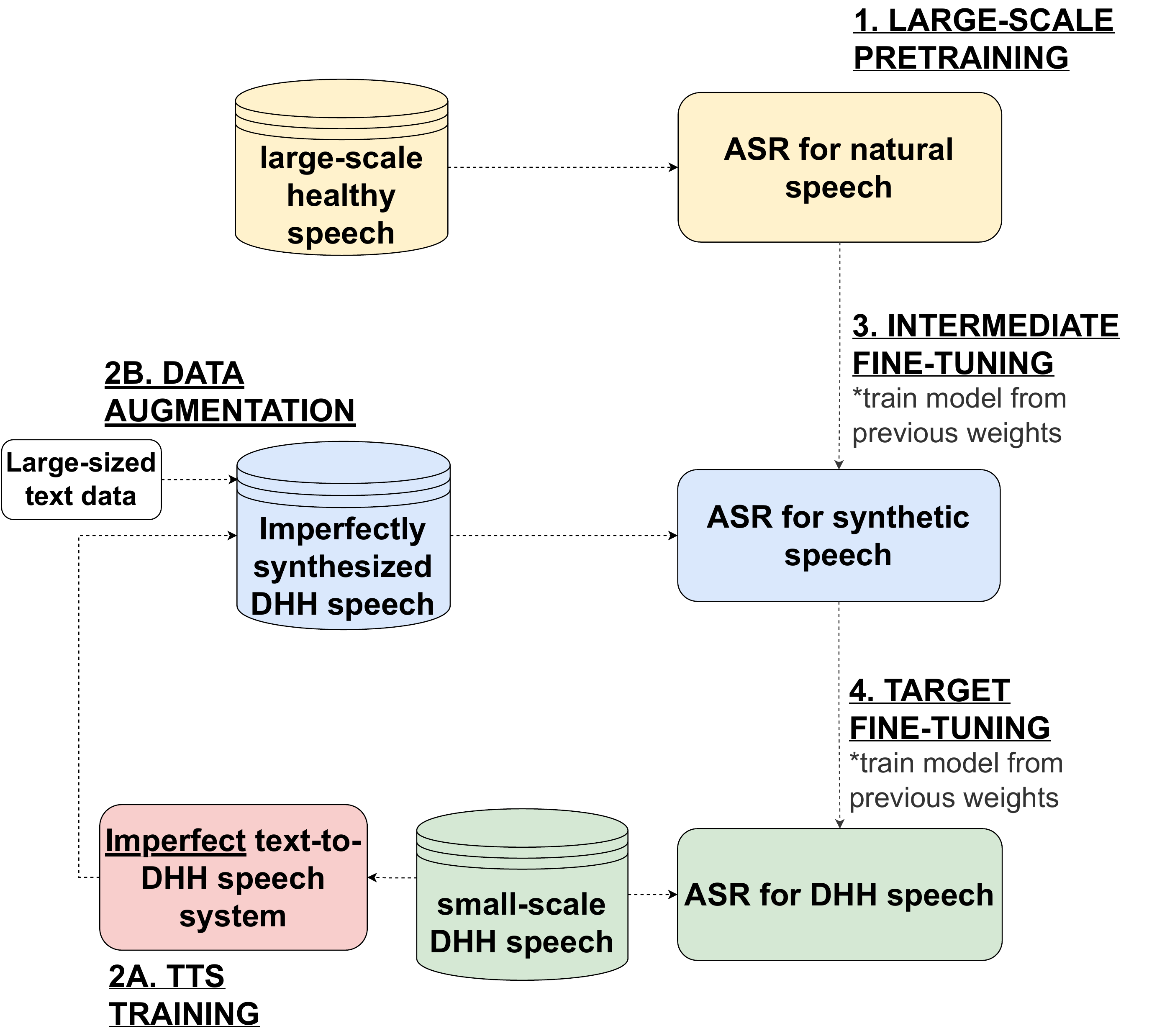}
    \caption{
    An overview of the ASR training process. We follow the setup described in \cite{el-intermediate-finetuning} and expand on the investigations of this work on deaf and hard-of-hearing speakers.
    }
    \label{fig:overview}
\end{figure}

\section{Methodology}
\subsection{Development of ASR model}
We expand on our previous work shown in \cite{el-intermediate-finetuning}, to verify the effectiveness of the intermediate fine-tuning task on other atypical speech datasets and with different training data sizes. Following this experiment, we primarily used setups based on pretraining on a large-scale typical speech dataset and subsequently fine-tuning on each target speaker. We then perform a thorough analysis of their behavior to understand how to develop efficient ASR models. To generate synthetic speech used for data augmentation, we use the same TTS setup used in that experiment. The TTS setup involves using a pretrained model and fine-tuning it on each of the speakers to create a larger training data. Then, text from a separate speech dataset is used to generate the larger speech dataset of each speaker. An overview of the ASR model training procedure is shown in Figure \ref{fig:overview}.

\subsection{DHH dataset overview}
The dataset we used was collected from four different DHH Japanese speakers in the comfort of their homes. To record the dataset, we asked them to utter the phrases from the Japanese speech corpus of Saruwatari Lab of the University of Tokyo (JSUT) dataset \cite{jsut}, a dataset spoken by a native female Japanese speaker. The dataset contains 7696 utterances and is commonly used for speech synthesis tasks. However, speakers were allowed to only record a subset of the dataset or skip some utterances if they felt uncomfortable saying some of the phrases. All datasets were recorded at a 44.1kHz sampling rate. A detailed look at the dataset can be found in Table \ref{tab:deaf_dataset}. Compared to commonly used atypical speech datasets such as UASpeech \cite{UASpeech} and TORGO \cite{TORGO}, our dataset is larger in total and contains a large number of utterances per speaker, making personalized ASR more possible. Moreover, compared to UASpeech \cite{UASpeech} which only contains word-level utterances, our dataset contains sentence-level utterances, making it a more realistic scenario for certain personalized ASR scenarios. With a larger dataset, we aim to analyze the behavior of the network when varying the training data sizes.

\begin{table}[t]
\centering
    \caption{An overview of the DHH speaker dataset collected.}
    \label{tab:deaf_dataset}
\begin{tabular}{llcc}
\toprule
\textbf{Speaker ID} & \textbf{Gender} & \textbf{No. of utterances} & \textbf{Total length (Hours)} \\
\midrule
S1 & Male & 5000 & 11.41 \\
S2 & Male & 4982 & 10.69 \\
S3 & Female & 902 & 1.64 \\
S4 & Female & 2493 & 4.31 \\
\midrule
Total & - & 13377 & 28.05 \\
\bottomrule
\end{tabular}
\end{table}

\section{Experiments}
\subsection{Experimental conditions}
To conduct the experiments, we used ESPnet \cite{espnet, espnet-2020}, an open-sourced speech processing toolkit. For the ASR model architecture, we used a Conformer \cite{conformer} encoder and a Transformer \cite{transformer} decoder due to their proven strength in ASR tasks \cite{espnet-conformer}. As for the TTS models, we used a Transformer \cite{transformer} encoder and decoder. Since the TTS models only convert the text to a mel-spectrogram, we used a pretrained model based on HiFiGAN \cite{hifigan} as the vocoder to convert mel-spectrograms to waveforms. The ASR and TTS pretrained models are made openly-available by ESPNet, while we also used a pretrained HiFiGAN model\footnote{\url{https://github.com/kan-bayashi/ParallelWaveGAN}} on the JSUT \cite{jsut} dataset. To pretrain the ASR model, we used the large-scale LaboroTV \cite{laborotv} dataset, which contains around 2000 hours of Japanese speech collected from various TV recordings. On the other hand, to pretrain the TTS model, we used the entire JSUT \cite{jsut} dataset.

To vary the training data sizes, we subset the datasets into different groups depending on the number of utterances available. More details on the number of utterances used are shown in Fig. \ref{fig:overview} as they differ for each speaker due to the total dataset size. We evaluated the trained model using a test set with 250 utterances for speakers S1, S2, and S4, and 50 utterances for speaker S3. Although the TTS models were trained and synthesized the utterances in 24 kHz, the synthesized audio would be downsampled to 16 kHz when used to train the ASR model. Thus, all audio used in the ASR models were in a 16 kHz sampling rate.

\subsection{Data augmentation setups}
We describe the different setups used in our experiment. We follow the intermediate fine-tuning (IF) setup detailed in \cite{el-intermediate-finetuning} and extend upon their findings.
\begin{itemize}
\item $\textbf{Direct}$: Directly using the out-of-the-box pretrained model and evaluating the performance without performing any fine-tuning.
\item $\textbf{Direct - speaker-independent}$: Fine-tuning the dataset with the other DHH speakers' audio data. This means that we do not use the target DHH speaker's audio data to fine-tune the model.
\item $\textbf{Naive fine-tuning}$: Simply fine-tuning the pretrained model on the target speaker without any extra procedures.
\item $\textbf{IF - TTS}$: Based on the intermediate fine-tuning setup in \cite{el-intermediate-finetuning}. We generate a larger dataset of the target DHH speaker using TTS, and use these during intermediate fine-tuning.
\item $\textbf{IF - TTS with text-swapping}$: Similar to the TTS setup, but randomly swapping the text labels within the dataset. Note that the text labels are only swapped during the intermediate fine-tuning stage. We further investigate this setup due to its reported success in \cite{el-intermediate-finetuning} but non-intuitive setup.
\item $\textbf{IF - speaker-independent}$: Instead of using synthetic speech, we use the real speech recordings from other DHH speakers during intermediate fine-tuning. This allows us to assess if the improvements in intermediate fine-tuning come from the speaker-dependent or speaker-independent characteristics of the data.
\end{itemize}
\begin{figure*}[ht!]
    \includegraphics[width=18cm]{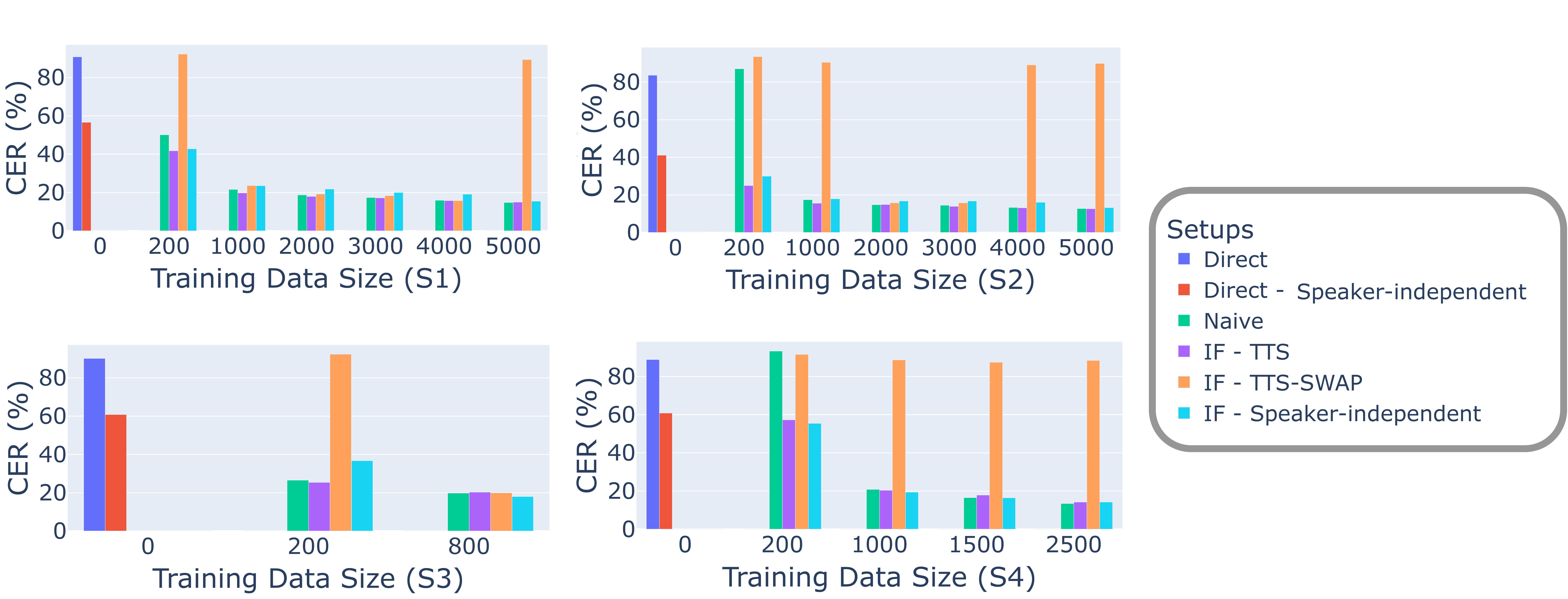}
    \caption{
    Character error rate (CER) of the different developed ASR systems for all speakers (S1, S2, S3, and S4). The x-axis indicates the amount of data used in the training and development sets.
    }
    \label{fig:s1}
\end{figure*}
\subsection{Evaluation metric}
To evaluate the ASR models, we use the character error rate (CER), as seen in Equation \ref{eq:cer}, a commonly used metric for Japanese ASR. Given an N number of characters and the number of wrongly predicted characters, which are categorized into three types, insertions (I), deletions (D), and substitutions (S), the CER can be calculated as the sum of the insertions, deletions, and substitutions divided by the number of characters, multiplied by 100. Thus, the lower the CER value, the better the performance of the ASR model.

\begin{equation}
    \label{eq:cer}
    \operatorname{CER} = \frac{\operatorname{I} + \operatorname{D} + \operatorname{S}}{\operatorname{N}} \operatorname{x} 100
\end{equation}

\section{Results and Analysis}
\label{sec:results}
\subsection{ASR model performance using out-of-the-box pretrained models}

We present the summary of our results in Figure \ref{fig:s1}. As displayed, in all cases where fine-tuning is not used, the ASR model always performed above 80\% CER as shown by the results of using the Direct setup. Thus, despite the ASR model being pretrained on a large-scale dataset, it still has difficulties adapting to the DHH speech, showing the difficulties faced by deaf speakers using out-of-the-box speech recognition devices. Aside from this, fine-tuning on speech data of other DHH speakers improves performance, but still not effective enough as shown in the Direct - speaker independent setup, proving the need for personalized ASR models. On the other hand, when fine-tuning the model on the available utterances (as shown by the Naive setup), the model's performance almost always improves, where the only exception is when using 200 utterances, similar to the findings in \cite{el-intermediate-finetuning}. However, using a larger dataset of the target speaker should be sufficient enough for the model to generalize. In all speakers, we find improvements of between 5\% to 30\% CER by increasing the number of utterances to 1000 from 200. Thus, we recommend researchers collect at least 1000 utterances as the ASR models can be already be effectively fine-tuned using this amount of data.

\begin{figure}[ht]
    \includegraphics[width=9cm]{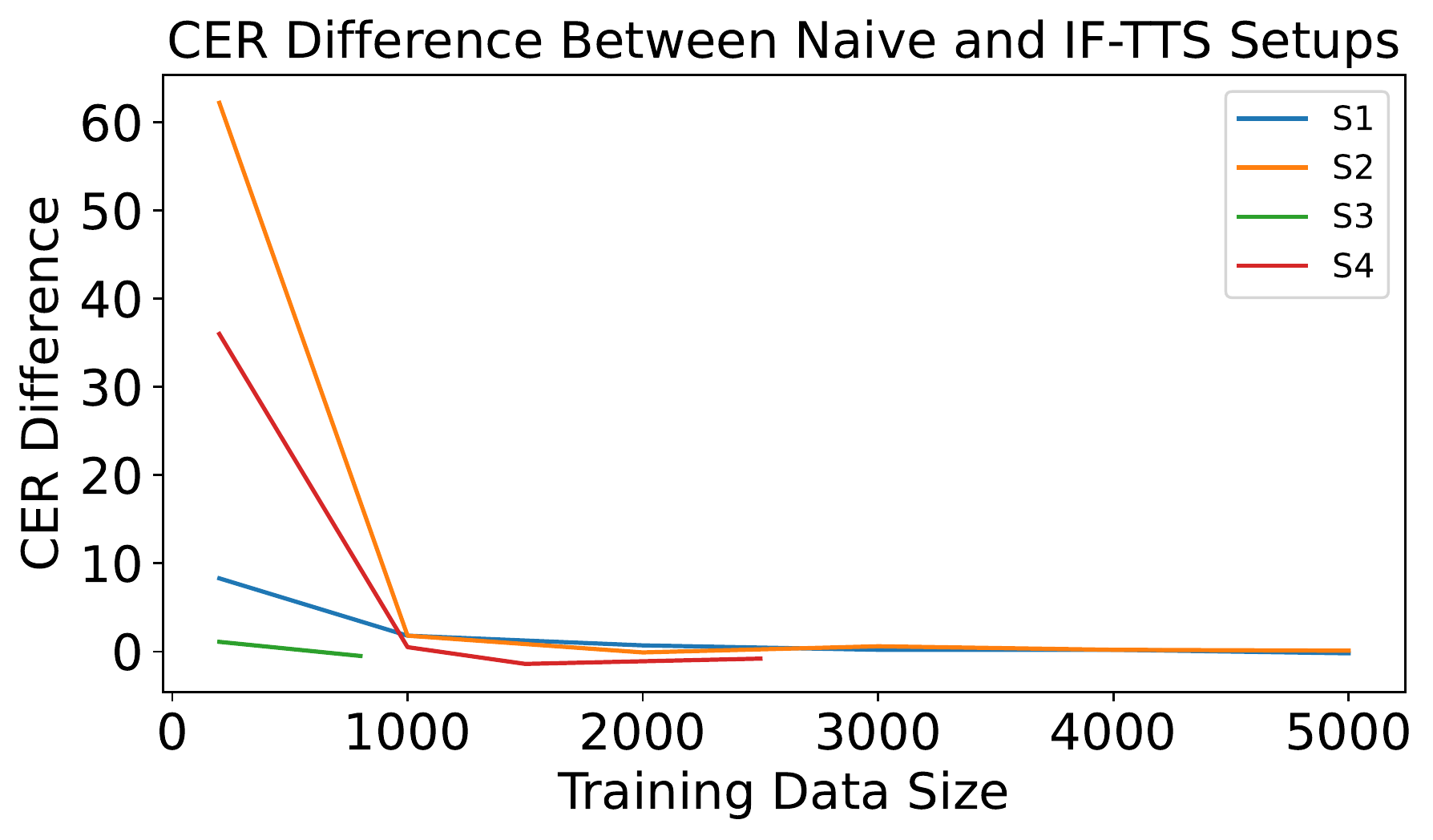}
    \caption{
    Difference of the character error rate (CER) of the Naive and IF-TTS setups across different training data sizes.
    }
    \label{fig:difference}
\end{figure}

\subsection{Extending investigations on intermediate fine-tuning}
We investigate whether our previously proposed data augmentation techniques for improving electrolaryngeal speech recognition also perform well on DHH speech. Using data augmentation techniques such as TTS to create a larger dataset of the target speaker and using intermediate fine-tuning, we see that this fine-tuning method can significantly improve performance in this low-resourced scenario, verifying that our method proposed in \cite{el-intermediate-finetuning} can also be transferred to DHH speech data. However, we observe that using intermediate fine-tuning became less effective as the training data size increased. Amongst the different setups, we find that the IF-TTS setup was the most stable and showed the most improvements in the low-resourced scenarios. Although the IF-TTS text-swapping setup was reported to be successful in \cite{el-intermediate-finetuning}, we do not see the same trend in our experiments as we observe its instability. Specifically, in cases where the training data is too low or too high, we observe a collapse and the ASR model fails. One hypothesis on why this is the case is because of the difference in speech features between electrolaryngeal speech and DHH speech. For electrolaryngeal speech, the speaker identity is disrupted, but the intelligibility is still there, as several phonemes can still be pronounced almost the same as a typical speaker. On the other hand, DHH speech contains the speaker identity but has unintelligible speech. Thus, since the intelligibility information between the pretraining data (LaboroTV) and the target data (electrolaryngeal speech) could be transferred and shared, the intermediate fine-tuning worked better on the electrolaryngeal dataset. On the other hand, in the case of our experiments, since the intelligibility information between the pretraining data (LaboroTV) and the target data (DHH speech) vastly differed, the information was not effectively shared. ASR models typically focus on the intelligibility information and disregard speaker identity information, which also explains why the text-swapping was not as effective in the case of DHH speakers. Finally, we observe that using data augmentations of the target speaker, even when the quality is low, was still always better than using real recordings of other DHH speakers as shown by the IF - speaker-independent setup results. 

As a final observation, although we verified the validity of the intermediate fine-tuning in the low-resource scenario, we still find that it cannot beat the performance when using real recordings of the target speaker as the training data. A detailed look on the comparison of the Naive and IF-TTS setups are shown in Fig \ref{fig:difference}. As seen in the figure, using imperfect synthetic speech only became successful in the 200 utterance setup as there was a larger CER difference between using the two setups; however, as more real data was collected and used during training, the effects of using imperfect synthetic speech became close to none. Thus, although we acknowledge the difficulty of obtaining training data from the target speaker, we urge researchers to collect at least 1000 utterances to easily develop efficient personalized ASR systems, and such that they would not need to do any additional procedures such as intermediate fine-tuning and speech synthesis to improve the ASR models. On the other hand, in cases where collecting data is difficult, researchers can still opt to use the intermediate fine-tuning procedure, as it has been verified to work even on other types of atypical speech.

\section{Conclusion}
We presented a thorough analysis of developing personalized ASR systems for DHH Japanese speakers. We provided a thorough extension of our previous work to verify its effectiveness by analyzing its behavior when varying the number of utterances in the training data and with a different atypical speech dataset. We collected a large-scale DHH dataset, totaling around 28.05 hours of four speakers. Our findings show the following observations: 1) the most efficient setup is to collect 1000 utterances from the target speaker, as we see the most performance boost compared to using just 200 utterances, 2) in cases where 1000 utterances are difficult to collect, data augmentation techniques such as using intermediate fine-tuning with synthetic speech is still an available option to improve performance, and finally, 3) we emphasized the importance of using personalized models, as using real recordings from the target speaker as training data greatly outperforms using synthetic speech or recordings from other speakers. Future work would be to improve the speech intelligibility of atypical speakers using voice conversion techniques such that they can easily use out-of-the-box ASR models.

\section*{Acknowledgment}
This work was partly supported by AMED under Grant Number JP21dk0310114, Japan, and a project, JPNP20006, commissioned by NEDO.
\bibliography{mybib}
\bibliographystyle{IEEEtran}

\end{document}